\documentstyle[prd,aps,preprint]{revtex}
\tighten

\def\S{{\cal S}}
\def\V{{\cal V}}
\def\be{\begin{equation}}
\def\ee{\end{equation}}
\def\bea{\begin{eqnarray}}
\def\eea{\end{eqnarray}}

\renewcommand\({\left(}
\renewcommand\){\right)}
\renewcommand\[{\left[}
\renewcommand\]{\right]}

\newcommand\eq[1]{Eq.~(\ref{#1})}

\newcommand\MeV{\,\mbox{MeV}}

\newcommand\lsim{\mathrel{\rlap{\lower4pt\hbox{\hskip1pt$\sim$}}
    \raise1pt\hbox{$<$}}}
\newcommand\gsim{\mathrel{\rlap{\lower4pt\hbox{\hskip1pt$\sim$}}
    \raise1pt\hbox{$>$}}}

\newcommand\diff{\mbox d}

\def\calp{{\cal P}}
\def\calr{{\cal R}}
\def\calv{{\cal V}}

\newcommand\bfk{{\bf k}}

\newcommand\bfx{{\bf x}}

\newcommand\sub[1]{_{\rm #1}}

\newcommand\cdm{\sub{cdm}}

\begin{document}
\preprint{astro-ph/0306498}

%
%
%

\title{Conserved cosmological perturbations}
\author{David H.\ Lyth$^1$ and David Wands$^2$}
\address{(1) Department of Physics, Lancaster University, Lancaster
LA1 4YB,~~~U.~K.}
\address{(2) Institute of Cosmology and Gravitation, University of
Portsmouth,\\Portsmouth PO1 2EG,~~~U.~K.}
\date{\today}
\maketitle
\begin{abstract}
  A conserved cosmological perturbation is associated with each
  quantity whose local evolution is determined entirely by the local
  expansion of the Universe. It may be defined as the appropriately
  normalised perturbation of the quantity, defined using a slicing of
  spacetime such that the expansion between slices is spatially
  homogeneous.  To first order, on super-horizon scales, the slicing
  with unperturbed intrinsic curvature has this property.  A general
  construction is given for conserved quantities, yielding the
  curvature perturbation $\zeta$ as well as more recently-considered
  conserved perturbations.  The construction 
 may be 
  extended to higher orders in perturbation theory and even into the
  non-perturbative regime.
\end{abstract}

\pacs{PACS numbers: 98.80.Cq\hfill astro-ph/0306498}



\section{Introduction}

Observation of the peak structure in the Cosmic Microwave
Background (CMB) anisotropy has now confirmed that cosmological
perturbations are present before the relevant scales enter the
horizon, with an almost flat (scale-invariant) spectrum
\cite{wmapspergel,wmappeiris,wmap}. 
The only known explanation for this state of
affairs is that the perturbations originate during an almost
exponential inflation, from the vacuum fluctuation of one or more
light scalar fields.\footnote {A related hypothesis
 replaces
inflation
  by an era of collapse (`pre-big-bang' \cite{prebigbang,BGGMV,LWC}
 or `ekpyrotic' 
\cite{ekpyrotic1,ekpyrotic2,pyro}), but there is
  so far no accepted theory of a  bounce and therefore no firm
  prediction 
from collapsing cosmologies. 
In particular, there is so far no accepted string-theoretic description of a
bounce \cite{lms}.}
 In the simplest case only one light field
 is responsible for the perturbations observed, either the
inflaton 
or some other 
field .

According to this explanation,  classical cosmological
perturbations first come into existence a few Hubble times after
horizon exit during inflation.  At that stage the situation is
very simple; each light field (defined as one with an effective
mass much less than the Hubble parameter $H$) has a Gaussian
perturbation with an almost flat spectrum, $(H/2\pi)^2$.  The
problem is to evolve this simple initial condition forward in time
to the primordial nucleosynthesis epoch, in the face of our
ignorance about the detailed evolution of the Universe before
nucleosynthesis.

Fortunately, scales of cosmological interest are still far outside
the horizon at nucleosynthesis. As a result there exist
perturbations which are under suitable conditions conserved, and
largely avoiding the need for more detailed information.  One of
these \cite{BST,Bardeen88,MS,WMLL} is the `curvature perturbation' $\zeta$,
which is associated with the perturbation in the total energy
density $\rho$.\footnote {The quantity $\zeta$ defines the curvature 
perturbation on spacetime slices of uniform energy density. As we
discuss in Section \ref{shear},  on  
super-horizon scales it is practically the same as $\calr$ which defines
the curvature perturbation on slices orthogonal to comoving worldlines.
The  latter quantity is the $\phi\sub m$ of \cite{bardeen80}.}
In the
usual case that  $\zeta$  originates from the perturbation in the
inflaton field, it is supposed to be conserved between the end of
inflation and the primordial era, and in the alternative curvaton
scenario 
\cite{lw,Moroi} (see also \cite{sylvia,lm})
$\zeta$  is supposed to be conserved after the curvaton
decays.\footnote
{An analogous scenario has been proposed
 in the pre big bang scenario \cite{Enqvist,bggv}.
In this scenario though,  the required scale-invariant curvaton
field perturbations will be generated only if the curvaton has a
non-trivial coupling and for particular initial
conditions~\cite{CEW,LWC}.}
 Recently, further conserved quantities $\zeta_i$ and
$\tilde\zeta_i$ have been considered, that are associated with the
perturbations in individual energy densities $\rho_i$ \cite{WMLL}
and number densities $n_i$ of conserved quantities \cite{luw}. The
conservation of the former is invoked in the curvaton scenario,
during the era when the curvaton field is oscillating and $\zeta $
is growing. The latter are invoked when considering possible
isocurvature components of the primordial density perturbation.

In this paper,  we present a unified treatment  of  the
conserved  quantities $\zeta$, $\zeta_i$ and $\tilde \zeta_i$,
which is more complete than anything that has been given before.
Taking the particular example of $\zeta$ as a starting point,
we  begin in Section \ref{deln} by showing how, to any order in
cosmological perturbation theory,  conserved quantities may be
constructed from perturbations that are defined on a spacetime
slicing of uniform integrated expansion. In Section
\ref{conserved} we generalise the construction and consider the
conserved quantities $\zeta_i$ and $\tilde\zeta_i$. In Section
\ref{uniformexp}, we show that in the usual case of first-order
perturbation theory, the spatially flat slicing is one of uniform
expansion if the shear of the worldlines is negligible. In Section
\ref{shear} we consider the comoving shear, and show that it is
expected to be negligible in the entire super-horizon regime. We
conclude in Section \ref{conclusion}. An Appendix mentions some
peripheral issues.

\section{Energy conservation and the curvature perturbation}

\label{deln}

In this section we explain the general principle which allows us
to construct conserved quantities. We focus on the particularly
important example of the curvature perturbation
$\zeta$~\cite{BST,Bardeen88,MS,WMLL}, after which it is clear how
other conserved perturbations may be constructed.

The curvature perturbation $\zeta$ is so called because it defines
the curvature perturbation on slices of uniform energy
density~\cite{WMLL}. Equivalently though, via the gauge
transformations of Section~\ref{uniformexp}, it defines the energy
density perturbation on spatially flat slices, according to the
formula~\cite{Bardeen88}
 \be
\zeta = \frac{\delta\rho}{3(\rho + P)}
 \label{zetadef0} \,.
 \ee
This definition is the one that we shall use.

Our starting point is the energy continuity equation. In an
unperturbed Friedmann-Robertson-Walker (FRW) universe
the continuity equation for the energy density $\rho$ takes the
form
 \be
 \dot \rho = -3H (\rho + P)
 \label{unpertenergy} \,,
 \ee
where $H$ is the Hubble expansion rate and $P$ is the pressure. In
the real perturbed Universe, the same equation
(\ref{unpertenergy}) still holds along each comoving worldline, so
long as the dot is taken to denote the derivative with respect to
the proper time, $\tau$, along the comoving worldline and we
define $H$ locally through the equation
 \be
 H \equiv \frac 13 \V \frac{\diff \V}{\diff \tau} \,,
 \ee
where $\V$ is an infinitesimal comoving volume. Equivalently, the
local continuity equation may be written as
 \be
 \V {\diff  \rho\over \diff  \V} = - (\rho + P) \,,
 \label{energy0}
 \,.
 \ee
or
 \begin{equation}
 \label{energy}
 {\diff \rho \over \diff  {N}}   = -3 ( \rho+ {P})
 \,.
 \end{equation}
where ${N}$ is the local logarithmic  integrated expansion
(the number of Hubble times)
defined as
\begin{equation}
 {N} \equiv \int  {H} d \tau
\label{nexp}
\,.
\end{equation}

Our crucial assumption now is that the pressure perturbation is
practically adiabatic.
This assumption means that the local pressure, $P$, is a practically unique
function of local energy density, $\rho$.
i.e.,
 \be
 P=\bar{P}(\rho) \,,
 \ee
where $\bar{P}$ is the same function for all worldlines. This
allows \eq{energy} to be integrated.
Setting $N=0$ on an initial spacetime slice
the integration gives $\rho$ as a unique function of the local
integrated expansion, $N$, upto an initial integration constant
 \be
 \rho = \bar\rho(N+\delta N)
 \label{adiabaticme} \,,
 \ee
where the integration constant for each worldline, $\delta N$, is
determined by the actual density on the initial hypersurface,
$\rho|_{N=0}=\bar\rho(\delta N)$.

Subsequent spacetime slices of fixed $N$ correspond to a {\em
uniform integrated expansion} slicing of the spacetime\footnote{
Note that this is {\em not} the same as the {\em uniform 
Hubble}
 slicing introduced by Bardeen~\cite{bardeen80,BST}
which  refers to the local expansion rate
of the normals.
},
meaning that the integrated expansion going from one slice to
another is spatially homogeneous. For linear perturbations about
an FRW cosmology, there is an infinity of such uniform-$N$
slicings, since we can start with any initial slice and propagate
it by calculating $N$ from that slice along each comoving
worldline.
In Sections \ref{uniformexp} and \ref{shear}, we show that on
super-horizon scales a particular uniform-$N$ slicing is the
uniform curvature slicing (i.e., the one with unperturbed
intrinsic scalar curvature).
In what follows we will restrict our attention to spatially flat
FRW models and will refer to this as the spatially flat slicing.

Now we come to the crucial point. When evaluating the density,
$\rho$, on any uniform-$N$ slicing, the 
perturbation $\delta N$ of the quantity appearing in \eq{adiabaticme} 
is time-independent, by
construction. This statement holds to any order in cosmological
perturbation theory
so long as one can construct 
a uniform-$N$ 
slicing along the comoving worldlines.
%

Writing $\delta N$ in terms of the density perturbation on
spatially flat slices, to first order, one finds the conserved
quantity
 \be
 \delta N =  \frac{\diff N}{\diff \rho} \delta \rho
 = \frac{\delta\rho}{\rho'(N)}
 = H\frac{\delta \rho}{\dot \rho}
 \label{zetaofn}
 \,.
 \ee
which is $-\zeta$ defined in \eq{zetadef0}.
%
%
%
To  arrive at the conserved quantity $\zeta$, we considered the
flat slicing.
Were we instead to use some other uniform-expansion slicing,
the conserved quantity defined by
the right hand side of \eq{zetaofn} would be different from 
$\zeta$, but it would be related to $\zeta$ by 
the gauge transformation \eq{delg}. Hence it would be conserved if and only
if $\zeta$ is conserved, and we lose no generality by
fixing the  choice of the uniform-expansion  slicing as the flat one.
%

The constancy of $\zeta$
(on sufficiently large scales and assuming that the pressure perturbation is adiabatic)
was obtained several years ago \cite{BST}
in the context of Einstein gravity. More recently, its constancy
under the same condition
was obtained directly from the local conservation
of energy \cite{WMLL} using a purely geometric argument equivalent
to the one that we have given. 
In the present paper, we are going to show in 
Sections  \ref{uniformexp} and \ref{shear} that in this context 
all super-horizon scales are `sufficiently large';
in other words, we will show that the flat slicing is a uniform-expansion
one on all super-horizon scales.

 It is worth noting 
that the conservation of $\zeta$ can hold in even more general
circumstances, because it comes from the generalised adiabatic
condition Eq.~(\ref{adiabaticme}) which may hold even if the
energy conservation equation~(\ref{energy}) fails. Thus, $\zeta$
will be conserved even if there is an additional source term ${Q}$
on the right-hand-side of in Eq.~({energy}), so long as ${Q}$ (the
energy transfer per Hubble time) is itself a unique function of
the local density for all worldlines, i.e., $
{Q}=\bar{Q}({\rho})$, as reported in Ref.~\cite{mwu}.

Going to second order, and again working on some uniform-$N$
slicing, the conserved quantity is
 \bea
 \delta N &=&  \frac{\diff N}{\diff \rho} \delta \rho
  +\frac12  \frac{\diff^2 N}{\diff \rho^2 }(\delta\rho)^2 \\
 &=&  \frac{\delta\rho}{\rho'}
 -\frac12  \frac{\rho''}{\rho'^3} (\delta \rho)^2
 \,.
 \eea
This second-order extension of the conserved quantity $\zeta$ has
not been given before. It will be useful in propagating forward
the evolution of second-order perturbations produced during
inflation \cite{ST,maldacena,toni} through the end of inflation
and relating them to observations. Also,  we note that
Sasaki and Tanaka \cite{ST} have shown that it is possible to use
a uniform-$N$ slicing to study non-linear field perturbations on
large scales during inflation.

\section{Other conserved quantities}

\label{conserved}


Generalising from the construction of $\zeta$, it is clear that
for any 
monotonically increasing or decreasing
quantity, satisfying a local conservation equation of the
form
 \be
 {\cal V}\frac{\partial f}{\partial \cal V} = y(f)
 \,,
 \ee
we can construct a conserved first-order perturbation
 \be
 X_f \equiv -H \frac{\delta f}{\dot f} \,,
 \ee
with $\delta f$ evaluated on some uniform-$N$ slicing which we
will take to be the spatially flat one. 
This construction gives
$\zeta\equiv X_\rho$ as a special case, and we shall now see how
it gives the other conserved quantities $\zeta_i$ and
$\tilde\zeta_i$ \cite{WMLL,luw}.

\subsection{Separately conserved energy densities}

Suppose the total energy density $\rho$ of the Universe is a sum
of components, $\rho_i$, each one of them either radiation or
matter, and with no energy transfer between the components. In
that case the pressure of each component is a unique function of
its energy ($P_i=\rho_i/3$ for radiation and $P_i=0$ for matter)
and each component satisfies its own
separate energy conservation equation (since there is no energy
transfer)
\footnote
{In this expression, 
 $\cal V$ is the  volume which is comoving  with the
flow of $\rho_i$. This is not strictly the same as the  volume
which is comoving 
with the flow of total energy density, but we are going to show
in Section \ref{uniformexp} that on super-horizon scales all comoving
volumes become equivalent.}
 \be
 {\cal V}\frac{\partial \rho_i}{\partial \cal V} = -(\rho_i+P_i)
 \,.
 \ee
As a result there are the separately conserved perturbations
 \bea
 \zeta_i  & \equiv & X_{\rho_i} \\
&=& -H \frac{\delta\rho_i}{\dot\rho_i} \\
&=& \frac13 \frac{\delta \rho_i}{\rho_i + P_i}
\,,
\label{zetaidef}
 \eea
where $\delta\rho_i$ is evaluated on the flat slicing. This is
another result of \cite{WMLL}.

One can express the total density perturbation, $\zeta$, as a
weighted sum of the separate $\zeta_i$;
 \be
 \zeta =
 \frac{\sum \dot\rho_i \zeta_i}{\sum \dot\rho_i}
 \label{zetaofzetai} \,.
 \ee
If the $\zeta_i$ are all equal, then $\zeta=\zeta_i$ which is
constant. Otherwise $\zeta$ may have some variation, determined by
the conserved isocurvature perturbations defined by
 \be
\S_{ij} \equiv 3 \( \zeta_i - \zeta_j \)
\,.
 \ee
The condition that the $\zeta_i$ are equal is just the adiabatic
condition, that all of the separate energy densities (and hence
the total pressure) are uniform on slices of uniform total energy
density.

There are two eras in the early Universe where
separately-conserved $\zeta_i$ have been invoked. One is the
comparatively late era, beginning when the temperature falls below
$1\MeV$ and ending when cosmological scales start to approach the
horizon.\footnote
{Recall that electron-positron annihilation and neutrino
decoupling both take place when the temperature is around
$1\MeV$.}


The energy density during this  era  has four components,
 \be
\rho = \rho\sub{CDM} + \rho_B + \rho_\nu + \rho_\gamma
\,,
 \ee
with the radiation (photons and neutrinos) dominating the matter
(Cold Dark Matter and baryonic matter). The values of the four
conserved quantities $\zeta\sub{CDM}$, $\zeta_B$, $\zeta_\nu$ and
$\zeta_\gamma$ determine the evolution of the entire set of
cosmological perturbations after horizon entry, and can therefore
be determined by observation. The three isocurvature perturbations
(conventionally defined relative to the photon density to be
$\S\cdm\equiv \S_{{\rm cdm}\gamma}$, $\S_B\equiv \S_{B\gamma}$ and
$\S_\nu\equiv \S_{\nu\gamma}$) are found by observation to be at
most of order $\zeta$~\cite{Durrer,Amendola,gl,wmappeiris}. Since
radiation dominates, one deduces from \eq{zetaofzetai} that
$\zeta$ is constant on large scales to high accuracy during this
era.

The other era, which occurs in the recently proposed curvaton
scenario \cite{lw,Moroi} (see also \cite{sylvia,lm}), 
is the era (after inflation, but
before primordial nucleosynthesis) when the massive curvaton
field, $\sigma$, oscillates ($P_\sigma=0$) in a radiation
background ($P_r=\rho_r/3$),
 \be
 \rho = \rho_\sigma + \rho\sub r \,.
 \ee
Here, $\zeta\sub r$ is supposed to be negligible so that the total
curvature perturbation, $\zeta$, is given by \eq{zetaofzetai} as
 \bea
 \zeta(t)  &=& \frac{3\rho_\sigma}{4\rho\sub r + 3\rho_\sigma}
  \zeta_\sigma  \label{zetacurv}
 \,.
 \eea
Well before the curvaton decays, the radiation is supposed to
dominate so that $\zeta$ grows like $\rho_\sigma/\rho\sub r\propto
a(t)$, providing an example where the total $\zeta$ is not
conserved on super-horizon scales after inflation.

\subsection{Conserved number densities}

If $n_i$ is a conserved number density, then $n_i$ is inversely proportional
to the  volume.
\footnote
{The volume should be the one comoving with the flow of the conserved
quantity, but as already stated we are going to show in
Sections \ref{uniformexp} and \ref{shear} that the choice of comoving
volume  is irrelevant  on super-horizon scales. This irrelevance is
assumed implicitly when the 
 conservation of $\tilde \zeta_i$ is discussed in
\cite{luw}.}
A conservation law of the form given in \eq{calv} is satisfied with
$y(f)=-f$ and $f=n_i$, leading to the conservation of the
first-order perturbation \cite{luw}
 \be
 \tilde\zeta_i  \equiv  X_{n_i} = \frac13 \frac{\delta n_i}{n_i}
 \,.
 \ee
%

These conserved quantities find an application \cite{luw,lw032} in
connection with the three isocurvature perturbations $\S\cdm$,
$\S_B$ and $\S_\nu$. 
Before  $T\sim 1\MeV$, these 
 quantities are not the appropriate ones to consider, because
the separate energy density perturbations $\zeta_i$ may 
vary with time or  be simply undefined. (The latter is the case
for $\zeta_B$ before the quark-hadron transition.)
One can however consider instead the number
density $n\cdm$ of Cold Dark Matter particles, the density $n_B$
of Baryon Number and the density $n_L$ of Lepton Number. Each of
these number densities corresponds to a conserved quantity after
some epoch, which may be regarded as the epoch when the quantity
originates. The corresponding perturbations $\tilde \zeta_i$ are
thus conserved, and after the temperature falls below $1\MeV$ they
determine the three isocurvature perturbations according to the
formulas \cite{luw}
 \bea
 \frac 13 \S\sub{cdm} &\equiv&  \tilde\zeta\sub{cdm} - \zeta \\
 \frac 13 \S_B &\equiv & \tilde\zeta_B - \zeta \\
 \frac 13 \S_\nu &\equiv & \frac{45}{7} \( \frac \xi \pi \)^2
 \( \tilde \zeta_L - \zeta \)
 \,.
 \eea
In the last formula, $\xi$ is the lepton asymmetry which must
satisfy the nucleosynthesis constraint $|\xi| < 0.07$. In this
way, the three isocurvature perturbations 
can be calculated (or shown to vanish) given a model of the
early Universe.

\section{Uniform expansion between flat slices}

\label{uniformexp}

The main goal of this section is to show that the 
local
expansion of the 
Universe between spatially flat slices is uniform on sufficiently
large scales where shear is negligible. This result is purely
geometric, making no reference to the theory of gravity. Our
treatment amplifies the original one in Ref.~\cite{WMLL}, and in
particular we show for the first time that the result is
independent of the 
%
spacetime threading that defines the
expansion. 

\subsection{The metric perturbation}

The unperturbed FRW metric and comoving coordinates are defined by
the line element
 \be
  \diff s^2 = -\diff t^2 +  a^2(t) \delta_{ij} \diff x^i\diff x^j \,,
 \label{FRW}
 \ee
corresponding to metric components $g_{00}=-1$,
$g_{0i}=0$ and $g_{ij} =\delta_{ij} a^2$.

We are interested in the perturbed spacetime that is our Universe,
which we assume can be described by linear perturbations about a
FRW geometry. To 
define the perturbations one has to choose a
coordinate system which reduces to \eq{FRW} in the limit where the
perturbations vanish. Such a coordinate system (gauge) defines a
time-slicing (the spatial hypersurfaces with constant time coordinate)
and a threading (the worldlines with constant space coordinates) of
the spacetime. Since the coordinate system is required  to
coincide with \eq{FRW} in the limit where the perturbations
vanish, the slicing and threading coincide with the unperturbed
ones in that limit. We shall take this requirement for granted when
referring to a `generic' slicing or threading.

Once the perturbations are defined, 
their  evolution to first order may be described using the unperturbed
coordinate system, and Fourier components with different  wavevectors $\bfk$
decouple. The super-horizon regime is the regime $aH/k\gg 1$.

In this paper we are interested in the scalar mode of the
perturbations in the metric (no gravitational waves or vorticity).
In a generic gauge the Fourier components of the metric
perturbation are  specified by functions $A$, $B$, $D$ and
$E$,\footnote {These are the quantities defined in \cite{LLbook},
and are equal respectively to the quantities $A$, $B^{(0)}$, $H_L$
and $H_T^{(0)}$ of Bardeen \cite{bardeen80}. The 
quantities  $\Delta
x$ and $\Delta t$ below are respectively the
 the $L$ and $aT$ of Bardeen, and the quantity $V$ in
\eq{DWshear} is the $v\sub S^{(0)}$ of Bardeen.}
 \bea
 \frac12\delta g_{00} &\equiv &  - A \label{a} \\
 2a^{-2} \delta g_{0i} &\equiv& - B_i \equiv  ik_i B \label{b}\\
 \frac12 a^{-2} \delta g_{ij} &\equiv&  \delta_{ij} D + P_{ij} E
 \label{de} \\
 &=& \delta_{ij} -\psi -\frac{k_i k_j}{k^2} E \label{psie}
 \,,
 \eea
where $P_{ij}$ projects out the traceless part;
 \be
P_{ij} \equiv -\frac{k_ik_j}{k^2} + \frac13 \delta_{ij}
\,,
 \ee
and
%
 \be
-\psi \equiv D + \frac13 E
\,.
\ee
Under the  coordinate transformation $t\to t +\Delta t$ and
$x^i \to x^i + \Delta x^i$, with the Fourier component of $\Delta x^i$
of the form
\be
\Delta x^i = -i\frac {k_i} k \Delta x \label{delxi}
\,,
\ee
the metric components transform according to
\bea
\Delta A &=&  - \dot {\Delta t}  \\
\Delta B &=& a \dot {\Delta x} + k\Delta t \\
\Delta D &=& -\frac k 3 \Delta x - H \Delta t \\
\Delta E &=&  k \Delta x \\
\Delta \psi &=& H \Delta t \label{delpsi}
 \eea
The perturbation $E$ can be eliminated by a transformation of the
spatial coordinates, while $\psi$ depends only on the slicing. The
latter defines the intrinsic scalar 3-curvature of the slicing,
and is called its curvature perturbation.

The perturbation in a generic quantity $g$ defined in the
unperturbed Universe, such as energy density, pressure or the
value of a scalar field, has the transformation
 \be
\Delta g = - \dot g \Delta t
\,.
\label{delg}
 \ee
Applied to the energy-density, this transformation along with
\eq{delpsi} leads to the gauge-invariant definition of the
`curvature perturbation' $\zeta$,
 \be
 \zeta = -\psi - H\frac{\delta \rho}{\dot\rho}
 \label{zetadef}
 \,.
 \ee
Evaluated on slices of uniform density it is indeed the curvature
perturbation, but evaluated on flat slices it specifies instead the
energy density perturbation through \eq{zetadef0}.

\subsection{Shear and the expansion rate}


A given threading of spacetime is associated with an expansion
$\theta$, a traceless symmetric shear $\sigma_{ij}$ and an
antisymmetric vorticity $\omega_{ij}$. At
a given spacetime point, in a locally-inertial
rest-frame, these quantities are given by the standard decomposition
\cite{LLbook}
\be
\partial_i w_j =
 {1\over3} \theta \delta_{ij} + \sigma_{ij} +
 \omega_{ij}
,.
 \ee
where $w_i$ is the three-velocity of the infinitesimally nearby
threads. The expansion $\theta$  gives the rate of increase with
respect to proper time $\tau$ of an infinitesimal volume $\calv$
expanding with the threads,
 \be
 \theta = \delta^{ij} \partial_i w_j  = \frac 1 {\V}
 \frac{\diff \V}{\diff \tau }
 \,.
 \ee
For the scalar perturbations that we are considering, the
 vorticity vanishes, and the
 shear  takes the form
 \be
 \sigma_{ij} = P_{ij} \sigma
 \,.
 \ee
We shall call $\sigma$ the shear as well.

In a given gauge, the metric perturbations determine
the  shear and expansion rate of the coordinate threads
according to the expressions \cite{ks84}
 \bea
\sigma & =&  \dot E \\
\delta \theta  &= &  3 \dot D   - 3 H A
\,.
 \eea
The second relation may be written as
 \be
 \delta \theta =  -3\dot\psi - \sigma - 3HA
 \label{expansion}
 \,.
 \ee

Going to a new threading corresponding to the transformation
\eq{delxi}, the change in the Fourier component of the local
velocity field of the threads  is
 \be
 \Delta w_i = \dot {\Delta x^i} = -i \frac{k_i}{k} \dot{\Delta x}
 \,,
  \label{delwi}
 \ee
The corresponding changes in the shear and the expansion are equal
and opposite,
 \be
 \Delta (\delta\theta) = - \Delta \sigma
  =  k \dot{\Delta x}
 \label{delthet} \,,
 \ee
while the perturbations $\psi$ and $A$ are unchanged. It follows
that \eq{expansion} is valid for any threading, not only for the
coordinate threading. (In \cite{WMLL}, \eq{expansion} was 
given for the special case of the threading normal to the
coordinate slicing.)

Following \cite{WMLL}  we consider, instead of $\theta$, the
expansion $\tilde\theta$ with respect to coordinate time. Its
perturbation is
 \be
 \delta\tilde\theta =  -3 \dot \psi -  \sigma
\label{delthettil}
\,.
 \ee
For the flat slicing ($\psi=0$) this becomes
 \be
\delta\tilde\theta  = -  \sigma \,.
\label{delthettil2}
 \ee
This result is true for any choice of the threading that defines
the expansion rate, and has been derived without any reference to
a theory of gravity. 

The expressions given so far are valid quite
generally. We are interested though in super-horizon scales.
On sufficiently large scales the shear must become negligible
compared with the Hubble parameter, so that we recover an
unperturbed FRW universe. It follows that $\delta \tilde\theta$ is
negligible on sufficiently large scales, or in other words that
the expansion between successive flat slices becomes unperturbed.
This means that on sufficiently large scales we can make the
approximation
 \be
 \tilde \theta(\bfx,t) = 3H(t) \,,
\label{tiltheta}
 \ee
where $H(t)$ is the usual unperturbed quantity and $\tilde\theta$
is the expansion with respect to coordinate time of a generic
threading. (Remember that we consider only threadings which
coincide with the unperturbed one in the limit of zero
perturbation.)

Using \eq{tiltheta}, we can combine the results of section II and
III to derive the following general result for cosmological
perturbations:
 \begin{quote}
Consider a monotonically increasing or decreasing quantity $f$,
defined in some region of  spacetime, and its first-order perturbation
$\delta f$ defined on the spatially flat slicing.
Consider also 
some 
threading of spacetime,
defining an infinitesimal volume element $\V$. If
%
%
$f$ satisfies
a local conservation equation of the form
 \be
 {\cal V}\frac{\diff  f}{\diff  \cal V} = y(f)
 \label{calv}
 \,,
 \ee
then the rate of change of the perturbation
 \be
 X_f \equiv -H\frac{\delta f}{\dot f}
 \label{defXg}
 \ee
is
 \be
\dot X_f = \frac13 \sigma
\label{xgdot}
\,
 \ee
where $\sigma$ is the shear of the threading. As a result, $X_f$
is conserved on sufficiently large scales, where the shear is
negligible.
\end{quote}


\subsection*{The variation of $\zeta$}
%
Taking the derivative of \eq{zetadef} and using the local
conservation of energy along comoving worldlines one finds
 \be
 \dot\zeta = - \frac H {\rho + P} \delta P\sub{nad}  - \sigma \,,
 \label{zetadot}
 \ee
where $\sigma$ is the shear of the comoving worldlines and the
non-adiabatic part of the pressure perturbation is
 \be
 \delta P\sub{nad} \equiv
 \delta P - \frac{\dot P}{\dot \rho} \delta \rho \,.
 \ee
This result was derived by essentially the above method in
\cite{WMLL}. It was first derived (by a different method, and actually
for the curvature perturbation $\calr$) in \cite{bardeen80},
 Eqs.~(5.19) to (5.21).
In the particular case that the Universe consists entirely of
matter and radiation, $\zeta$ is given by \eq{zetaofzetai}. One
easily checks \cite{luw} that this expression is compatible with
\eq{zetadot}.

\section{Shear on super-horizon scales}

\label{shear}

According to \eq{xgdot}, the quantity $X_f$ is conserved on scales
which are sufficiently large that $\sigma$ is negligible. 
In this section we argue that any super-horizon scale is `sufficiently
large' in this context. To be more precise, we argue that 
 the shear satisfies
\be
|\sigma|/H \ll (k/aH)
\label{sigbound}
\,,
\ee
 which ensures that in one Hubble
time the change in $X_f$ is less than $k/aH$.\footnote
{On the left  hand side of this expression, $\sigma$ is  
the typical magnitude of the shear on scale $k$, defined for instance as
\footnote{LLbook}  the square root of its spectrum $\calp_\sigma(k)$.}

In making the argument, we  shall invoke
the Einstein field equations. This is not much of a restriction as
we do not necessarily have to specify any physical origin for the
stress-energy tensor, but simply equate it with a fixed multiple
of the Einstein tensor derived from the metric. This leads to a
purely geometrical definition of, e.g., the comoving density
perturbation, which need have nothing to do with the motion of
particles. However at relatively late cosmic times, it may be safe
to assume that the Einstein field equations are satisfied with the
stress-energy tensor related to particle physics content in the
usual way.

The other assumption we make is that anisotropic stress is negligible
on super-horizon scales. As noted by Bardeen in 1980 \cite{bardeen80},
significant  anisotropic stress on super-horizon scales  would 
generate shear and the curvature perturbation, but there is no known
mechanism for generating such stress. 

We first note that any spatial gauge transformation, \eq{delxi}
corresponds to a change in the local physical velocity $\Delta v_i \equiv
a\Delta w_i$ (\eq{delwi}). This generates a change in the shear 
$|\Delta\sigma|/H = v k/aH$). Since we are dealing with small perturbations,
$v\ll 1$ so that 
$|\Delta\sigma|/H \ll  k/aH$. 

It follows that we need only establish \eq{sigbound} for the shear of
the comoving threading, which $\sigma$ shall denote from now on. 
Generalising the discussion of Bardeen \cite{bardeen80} to include
the case where $P/\rho$ may vary, we 
 shall show that in fact
\be
|\sigma|/H \ll (k/aH)^2
\label{sigbound2}
\,.
\ee

The comoving shear is related to 
 two commonly used gauge-invariant variables,
namely the  curvature perturbation  $-\cal R$ of slices orthogonal to
comoving worldlines (comoving slices)
and the curvature perturbation $\Phi$ of  zero-shear hypersurfaces
(the Bardeen potential):
\footnote
{We are defining $-\calr\equiv \psi$ with the right hand side evaluated
on comoving slices, which corresponds to established conventions.}
\be
 \frac{\sigma}H =
 \frac{k}{aH}V =
 \( \frac{k}{aH} \)^2 \( {\cal R} + \Phi \)
\,. \label{DWshear} \ee
(The quantity $V$ defines the velocity $v_i$
of the comoving worldlines relative to
the the zero-shear threading, through the relation
$v_i = -i(k_i/k) V$ \cite{LLbook}.)
The curvature perturbation $\cal R$
is closely related to the
curvature of uniform-density slices, $\zeta$:
\be
{\cal R} = \zeta - \frac{H \delta\rho\sub{com} }{\dot \rho}
\,,
\ee
where the subscript `com' denotes the comoving slicing.

The combined energy and momentum constraints of Einstein's
equations relate the comoving density perturbation to the
Bardeen potential
\be
 \frac{H\delta\rho\sub{com}}{\dot\rho}
 =
 \frac 2{9(1+w)} \( \frac k {aH} \)^2 \Phi \,,
 \label{defrhocom}
\ee
where $w\equiv P/\rho$.
Thus we can rewrite Eq.~(\ref{DWshear}) for the comoving shear in
terms of $\zeta$ and the Bardeen potential, giving
\be
 \frac{\sigma}H =
 \( \frac{k}{aH} \)^2 \zeta + \( \frac{k}{aH} \)^2 \[ 1 - {2\over9(1+w)} \( \frac{k}{aH} \)^2 \] \Phi
 \,. \label{DWshear2}
\ee
%

Equation~(\ref{DWshear2}) ensures that the comoving shear will
be small ($\sigma/H\ll \zeta$) on 
super-horizon scales
so long as the Bardeen potential, $\Phi$, remains of the same
order as $\zeta$.
Notice also that if the $\Phi$ remains finite on large scales,
then the comoving density perturbation \eq{defrhocom} also
vanishes on large scales, and ${\cal R}$ and $\zeta$ are equal
(and constant) on large scales.
However, it has been argued \cite{bardeen80} that  cosmological
perturbation theory can still be valid even if certain curvature
perturbations, in particular $\Phi$, become formally bigger than
one.

The Bardeen potential is not uniquely determined by the value of
$\zeta$, but the Einstein's equations give a first-order evolution
equation \cite{LLbook} 
(in the absence of anisotropic stress \cite{bardeen80})
\be
 H^{-1} \dot\Phi + \[ \frac{5 + 3w}{2} -
  \frac 1 3 \( \frac k {aH} \)^2 \] \Phi
 =  - \frac 3 2 \(1 + w \)  \zeta \,. \label{dotphi}
\ee
During conventional slow-roll inflation (with $w\sim-1$ and
$\dot\Phi\ll H\Phi$), the Bardeen potential is indeed small,
$\Phi\sim -3(1+w)\zeta/2$ on super-horizon scales. But
we wish to eliminate the possibility that
the Bardeen potential subsequently becomes large on super-horizon scales.

\subsubsection{Adiabatic perturbations}

For strictly adiabatic matter perturbations, with $\delta P_{\rm
  nad}=0$, Eqs.~(\ref{zetadot}), (\ref{DWshear2}) and~(\ref{dotphi})
yield coupled first-order equations for the evolution of $\zeta$ and
$\Phi$, which to lowest order in $k/aH$ gives
\bea
 H^{-1} \dot\zeta &\simeq& \( \frac k {aH} \)^2 \Phi \,,
\label{zetadot2}
 \\
 H^{-1} \dot\Phi + \frac{5 + 3w}{2} \Phi &=&  -\frac32 \(1 + w \)
 \zeta\,,
\eea
which yield two independent long-wavelength solutions, which are
represented by
\bea
 \zeta &\simeq& C_+ \,,\\
 \Phi &\simeq& C_- e^{-(5+3\tilde{w})N/2} \,,
\eea
where $\tilde{w}N=\int wdN$.
The first of these solutions, with constant $\zeta$ on large
scales, remains the ``growing mode'' solution so long as
\be
 H^{-1} \dot\zeta
 \propto C_- k^2 e^{3(2w-\tilde{w}-1)N/2} \,,
\ee
for the ``decaying mode'' on large scales, approaches zero. This
is always true in an expanding universe ($N\to+\infty$) so long as
$w\to w_\infty<1$, i.e., $P<\rho$. This is easily interpreted as
the condition for the decay of the shear relative to the Hubble
rate ($\sigma/H$) in an expanding universe.

Using these same equations, we can understand the  super-horizon 
evolution  of the 
shear and the curvature perturbations in a collapsing universe
($N\to-\infty$).
For $w<1$,  the shear grows relative to the Hubble rate,
  and $\zeta$ does not
remain constant.
The critical case $w=1$ (maximally stiff fluid) occurs if the energy density
is dominated by scalar fields with negligible potential.
This is supposed to happen  in the pre big bang scenario, 
and in the late stages of the second version of the ekpyrotic scenario
\cite{ekpyrotic2} where the bounce is supposed to be singular from the
four-dimensional viewpoint. For $w=1$, 
 $\zeta$ on super-horizon scales grows logarithmically with respect
to cosmic time
and has a strongly scale-dependent spectrum \cite{BGGMV}.
(It is however \cite{BGGMV}
still small at the string 
epoch, which in the pre big bang scenario is supposed to be the bounce
epoch.)

In the first version of the ekpyrotic scenario \cite{ekpyrotic1},
where the bounce is supposed to be non-singular from the
four-dimensional viewpoint, collapse is driven by a scalar field with
a steep negative potential which violates the dominant energy
condition and gives $w\gg1$.  The same is supposed to happen in
the second version of the ekpyrotic scenario \cite{ekpyrotic2} at
early times.  In these  cases, the shear rapidly decreases
\cite{Heard,gratton} and $\zeta$ is
constant on large scales, with a strongly scale-dependent spectrum
$\calp_\zeta^\frac12 \propto k^{-2}$.
The Bardeen potential $\Phi$, related to $\zeta$ by \eq{zetadot2},
grows rapidly and has a flat spectrum \cite{bf,ekpyrotic2,Durrer2} 
$\calp_\Phi^\frac12 \propto k^0$.
But equation~(\ref{xgdot}) shows that it is only the comoving shear
that affects $\zeta$, and the shear is related to spatial gradients of
the Bardeen potential, Eq.~(\ref{DWshear2}). A scale-invariant Bardeen
potential ($\Phi\propto k^0$) corresponds to a strongly tilted blue
spectrum for the shear ($\sigma\propto k^2$).

\subsubsection{Non-adiabatic perturbations}

If we relax the requirement 
that the matter perturbations are adiabatic 
(but still assuming no anisotropic stress) 
then we no longer have a
closed system of equations for $\zeta$ and $\Phi$. However, we can
still estimate $\Phi$ in the long-wavelength regime given $\zeta$ and
integrating
\bea
\frac23 H^{-1} \dot\Phi + \frac{5 + 3w}{3} \Phi
& \simeq & -(1+w)\zeta
\,.
\label{phidot}
\eea
We also need an initial condition a few Hubble times after
horizon exit during slow-roll inflation.
Starting with the vacuum fluctuation, direct calculation shows that
$\zeta$ at this stage is  either practically constant
(single-field inflation) or only varying slowly on the Hubble
timescale (multi-field inflation). Through
\eq{phidot} this gives  $\Phi\simeq -(3/2)(1+w)\zeta$,
and hence $|\Phi|\ll |\zeta|$.

We are now going to argue that \eq{phidot} will keep $|\Phi|\lsim |\zeta|$
throughout the super-horizon era.
A rough argument is the following. Suppose that
instead  $|\Phi|\gg |\zeta|$
in some super-horizon regime. Then \eq{phidot} becomes
\be
\frac23 H^{-1} (\ln\Phi)\dot{} \simeq - \frac{5 + 3w(t)}{3}  < -(2/3)
\,,
\ee
where we used the energy condition $w>-1$ which is always satisfied in
scalar field theory with a positive kinetic energy.  This equation
shows that $|\Phi|$ would always be decreasing in any expanding
universe where $|\Phi|\gg |\zeta|$, suggesting that such a regime
cannot actually be reached starting from the initial condition
$|\Phi|\ll |\zeta|$.




A more direct argument is to integrate \eq{phidot}, giving
\be
F\Phi = -\frac32 \int^{\ln a}_{\ln a_1} \(1+w(a')\) F(a')
\zeta(a') \diff (\ln a')
\,,
\label{phiint}
\ee
where
\be
\ln F = \int^{\ln a}_{\ln a_1} \(
 \frac{5+3w(a')}{2}
   \)
 \diff (\ln a')
\,.
\ee
Assuming that $\zeta$ is never very much bigger than its primordial value,
this will give $|\Phi|\lsim |\zeta|$ for any reasonable behaviour of $w(a)$.

\section{Conclusions}

\label{conclusion}

In this paper we have shown how local conservation laws (e.g.,
energy conservation or baryon number conservation) can lead to
conserved perturbations in cosmology. Whenever we have a local
continuity equation of the form given in equation (\ref{calv}),
then we can construct a cosmological perturbation which is
conserved after uniform expansion along comoving worldlines.

In particular we have shown that in linear perturbation theory the
integrated expansion along comoving worldlines between spatially
flat slices is just given by the comoving shear. Thus on
sufficiently large scales (where the shear is negligible) the
quantity $X_f$ defined in Eq.~(\ref{defXg}) derived from the
conservation equation~(\ref{calv}) is conserved. The choice of
spatially flat slices gives a gauge-invariant definition of the
conserved quantity. This is a purely geometrical result, whose
derivation does not require any gravitational field equations. We
only require the gravitational field equations in order to
estimate the actual comoving shear,
finding it to be  negligible on super-horizon scales.

The best known example is the curvature perturbation $\zeta\equiv
X_\rho$,  which specifies the total density perturbation on
spatially flat slices or equivalently the curvature perturbation on
uniform-density slices. $\zeta$ is constant on sufficiently large
scales (where the comoving shear is negligible) for adiabatic
density perturbations, for which the local pressure is a unique
function of the local density and hence the total energy
conservation is of the form required in equation~(\ref{calv}). 
We have shown that on super-horizon scales,
$\zeta$ coincides with the comoving curvature
perturbation.

It is also possible to construct other perturbed quantities, such
as the separate curvature perturbation $\zeta_i\equiv X_{\rho_i}$
for any perfect fluid whose energy is separately
conserved~\cite{WMLL}, or $\tilde\zeta_i\equiv X_{n_i}$, for any
conserved number density, $n_i$, obeying a local conservation
equation of the form $\dot{n}_i=-3n_i$ \cite{luw}.

We also give an expression for the conserved quantity to
second-order in the density perturbation which may be employed in
calculations of the primordial non-Gaussianity of density
perturbations produced from inflation.

\acknowledgments We are grateful to Marco Bruni and Takahiro
Tanaka for useful discussions about non-linear perturbations. This
work was supported in part by
PPARC grants PPA/G/S/1999/00138, PPA/G/S/2000/00115 and
PGA/G/O/2000/00466.  DW is
supported by the Royal Society.

\section*{Appendix}

\subsection*{The Sasaki-Stewart expression for $\zeta$}

On super-horizon scales, \eq{delthettil} becomes
\be
\delta\tilde \theta = -3\dot \psi
\,.
\ee
This is valid in any gauge. Choose now a gauge whose slicing is flat 
 at time $t_1$, and uniform-density at time $t$. Integrating from 
$t_1$ to $t$,  and using $\zeta=-\psi$ on the slice at $t$,
we find that on super-horizon scales
\be
\zeta(\bfx,t) = \frac13 \int^t_{t_1} \tilde\theta \diff t = 
\delta N\sub{SS}(\bfx,t)
\ee
where $N\sub{SS}(\bfx,t_1,t)$ 
is the integrated expansion from the flat slice at 
time $t_1$, to the uniform-density slice at time $t$. This is the expression
of Sasaki and Stewart \cite{ss}, 
used by them to calculate the curvature  perturbation at the end of 
multi-field inflation. It is practically independent of the 
threading that defines the expansion, by virtue of the fact that
we are dealing  with super-horizon scales. 

Our expression \eq{zetaofn} reads $\zeta= -\delta N$. In contrast with the
Sasaki-Stewart expression, this one is valid only during an era when
$\zeta$ is constant, corresponding to an adiabatic  pressure perturbation.
To understand the relation with the Sasaki-Stewart
expression, we can integrate \eq{energy} from  a uniform-density slice
at time $t$ to a flat slice at time $t_1$, both times being within
the era when the pressure perturbation is adiabatic. Using 
$3\zeta=\delta\rho/(\rho+P)$ on the slice at $t_1$ we  get the time-independent
result
\be
\zeta = -\delta N
\,,
\ee
where $N=-N\sub{SS}$ is the integrated expansion {\em from} $t$ {\em to}
$t_1$. We see that the `integration constant' $\delta N$ introduced in
Section \ref{deln} can be interpreted as a perturbation in the integrated
expansion between two  slices, and that it is equal (as it must be)
 to $-\delta N\sub{SS}$.

\subsection*{Uniform  Hubble parameter on comoving slices}

In a given gauge, the perturbation in the expansion 
$\tilde\theta$ with respect to coordinate time
is given by 
\eq{delthettil}, which for the flat slicing becomes
\eq{delthettil2}, 
\be
\delta\tilde\theta = -\sigma
\label{a1}
\,.
\ee
On super-horizon scales this gives  $|\delta\tilde \theta|/H \ll 1$,
valid for any threading. In other words, the expansion with respect to 
coordinate time is practically unperturbed on flat slices.

We could instead consider the perturbation in the expansion with respect
to proper time, given by \eq{expansion}. On the  
the comoving slicing
one has 
in the absence of anisotropic stress
(Eqs.~(5.19) to (5.21) of Bardeen \cite{bardeen80};  see also
\cite{lythmuk,LLbook})
\be
\dot\calr\equiv -\dot\psi\sub{com} = H A\sub{com}
 \(= - H\frac{\delta P\sub{com}}{\rho + P} \)
\,,
\ee
and therefore
\be
(\delta\theta)\sub{com} = -\sigma
\,.
\ee
Like \eq{a1}, this expression is valid for any choice of the threading that
defines the expansion, and on super-horizon scales it gives
$|\delta\theta|/H \ll 1$ practically independently of the threading.
In other words, the expansion with respect to proper time is practically
unperturbed on comoving slices, for any choice of the threading.
In  particular the comoving expansion 
is practically unperturbed on such slices,
$\delta H/H \ll 1$.

We argued in Section V that  $\Phi$ is  
finite on super-horizon scales (in fact, that it is most  of
 order the curvature perturbation), and from \eq{defrhocom}
this implies that on comoving slices the density contrast on super-horizon
scales is practically zero,
\be
|\delta\rho\sub{com}/\rho| \ll 1
\,.
\ee

To summarise: on comoving slices,
  the perturbations in the
locally-defined Hubble parameter and in the energy density
are both negligible  in the super-horizon regime.
The only significant perturbations on comoving slices 
are therefore  curvature perturbation
$\calr$, and the pressure perturbation if it is not adiabatic.
The statements of the previous paragraph remain true if we replace
the comoving slicing  by the uniform-density slicing, since we argued in 
Section \ref{shear} that these slicings  practically
coincide on super-horizon scales.

\subsection*{The adiabatic condition on the pressure perturbation}

In the text we defined the adiabatic condition on the
 pressure perturbation as the condition that the local 
pressure 
is a practically unique function of the local energy density. 
Taking it to be absolutely unique, we obtain the familiar adiabatic
condition 
\be
\delta P = (\dot P/\dot \rho) \delta \rho
\label{strict}
\,.
\ee
However, on the  comoving slicing where $\delta \rho$ is anomalously
small, and on the uniform-density  slicing  where it vanishes,  it is too
strong to require that this expression  is valid; there is no reason
why the slices of uniform pressure should exactly coincide with the 
slices of uniform energy density even if the local pressure is a
`practically' unique function of the local energy density.

An example is provided by single-field slow-roll inflation. During 
slow-roll the locally-defined inflaton field is a practically unique 
function of proper time,  $\phi(\tau)$, up to the choice of origin for $\tau$.
On super-horizon scales, where spatial gradients are practically negligible,
this  gives practically unique  functions  $\rho(\tau)$  and $P(\tau)$,
making $P$ a practically unique function of $\rho$. In other words, the
 adiabatic condition for the pressure perturbation is satisfied 
on super-horizon scales during 
single-field slow-roll inflation (and even afterwards provided that 
no other field plays a significant role). However, on comoving slices
the potential $V(\phi)$ is uniform, and as a result 
\be
\delta P\sub{com} = \delta\rho\sub{com}
\,.
\label{pcom}
\ee
This is not  in accordance with the strict definition \eq{strict}
of an isocurvature pressure perturbation.
In particular, during slow-roll
inflaton $\rho\simeq -P (\simeq V)$ which means that the 
 adiabatic condition for the pressure perturbation is
\be
\delta P \simeq - \delta \rho
\,.
\ee
On a generic slicing this is well-satisfied, but on the comoving
slicing it is at variance with \eq{pcom}. 
All that matters, though, is that both the  pressure perturbation
and the energy density perturbation are both 
very small on the comoving slices. Or, to put it differently, that the
pressure perturbation is very small on  uniform-density slices.
This is enough to ensure that the local 
pressure is a practically unique function of the energy density,
leading to the conclusion that  $\zeta$ is constant
during single-field inflation.


\end{document}